\documentclass[12pt]{article}

 \newread\testifexists
 \def\GetIfExists #1 {\immediate\openin\testifexists=#1
     \ifeof\testifexists\immediate\closein\testifexists\else
     \immediate\closein\testifexists\input #1\fi}

 \usepackage{gthstyle}\usepackage{amsfonts}
 \usepackage{amssymb}
 \usepackage{graphicx} \usepackage{epstopdf}
 \def\Bbb#1{\setbox0=\hbox{$\tt #1$}  \copy0\kern-\wd0\kern .1em\copy0}
 \def\bbf#1{\setbox0=\hbox{$#1$} \kern-.025em\copy0\kern-\wd0
         \kern.05em\copy0\kern-\wd0 \kern-.025em\raise.0433em\box0}
 \immediate\openin\testifexists=a
     \ifeof\testifexists\immediate\closein\testifexists\else
     \immediate\closein\testifexists\input a\fimssym.def  % for \Bbb A - Z %%

                     \newcommand{\fn}{\footnote}
              \newcommand{\nm}{\nonumber}
 \newcommand{\be}{\begin{eqnarray}}             \newcommand{\ee}{\end{eqnarray}}
 \newcommand{\bi}[1]{\begin{itemize}\item[#1]}         \newcommand{\itm}[1]{\item[#1]}  \newcommand{\ei}{\end{itemize}}
 \newcommand{\eqn}[1]{(\ref{#1})}

 \newcommand{\crlb}[1]{\label{#1}\\[2pt]}
 \newcommand{\eela}[1]{\quad\hbox{\scriptsize{#1}}\label{#1}\end{eqnarray}}
 \newcommand{\eelb}[1]{\label{#1}\end{eqnarray}}
 
 \newcommand{\newsecb}[2]{\section{#1}\label{#2}\setcounter{equation}{0}}

 \newcommand{\nolabels} {\def\eel{\eelb} \def\crl{\crlb} \def\newsecl{\newsecb}\def\bibiteml{\bibitem}\def\citel{\cite}}

\newcommand\publishversion{\nolabels\setlength{\textheight}{9in}\setlength{\oddsidemargin}{0in}
    \setlength{\textwidth}{6.3in}\setlength{\topmargin}{-0.1in}}

 \def\a{\alpha}               
 \def\d{\delta}      \def\D{\Delta}   
 \def\k{\kappa}      \def\l{\lambda} \def\L{\Lambda}     
             \def\vv{\varphi}    
 \def\j{\psi}

       \def\W{\Omega}  

    \def\LL{{\mathcal L}}       
 \def\pa{\partial} \def\ra{\rightarrow} 
  
 \def\dd{{\rm d}}  \def\bra{\langle}   \def\ket{\rangle}

 \def\qu{\ {\buildrel {\displaystyle ?} \over =}\ }  
 
 \def\iss{\ =\ }

 \def\fract#1#2{{\textstyle{#1\over#2}}}
 \def\ffract#1#2{\raise .2 em\hbox{$\scriptstyle#1$}\kern-.3em/
                 \kern-.2em\lower .15 em \hbox{$\scriptstyle#2$}}
 
 \def\half{\fract12} \def\quart{\fract14} \def\halff{\ffract12}
 
 \def\part#1#2{{\partial#1\over\partial#2}}
 
 \def\Ex#1{\E^{\textstyle#1}}

			\def\E{\epsilon}		%\def\E{\hbox{\scriptsize{E}}}  	

\def\ds{\displaystyle}	
      
\def\intt{\mathrm{int}}

\publishversion

\begin{document} \begin{titlepage}

\title{\normalsize \hfill ITP-UU-12/18  \\ \hfill SPIN-12/16
\vskip 20mm \Large\bf Duality between a deterministic cellular automaton and a bosonic quantum field theory in 1+1 dimensions}

\author{Gerard 't~Hooft}
\date{\normalsize Institute for Theoretical Physics \\
Utrecht University \\ and
\medskip \\ Spinoza Institute \\ Postbox 80.195 \\ 3508 TD Utrecht, the Netherlands \smallskip \\
e-mail: \tt g.thooft@uu.nl \\ internet: \tt
http://www.phys.uu.nl/\~{}thooft/}

\maketitle

\begin{quotation} \noindent {\large\bf Abstract } \medskip \\
Methods developed in a previous paper are employed to define an exact correspondence between the states of a deterministic cellular automaton in
1+1 dimensions and those of a bosonic quantum field theory. The result may be used to argue that quantum field theories may be much closer related to deterministic automata than what is usually thought possible.\end{quotation}

\vfill \flushleft{May 21, 2012}

\end{titlepage}
\eject

%%%%%%%%%%%%%%%%%%%%%%%%%%%%%%%%%%%%%%%%%%%%%%%%%%%%

\newsecl{Introduction}{intro}

The idea that deterministic theories and quantum theories may be very closely related, has been advocated by this author for a number of years\cite{GtHCA}. To understand this relation, some issues needed to be further clarified. It has always been obvious that deterministic, time reversible systems can be cast in a quantum mechanical formalism, where the evolution is described by a hamiltonian that in no fundamental way differs from other, obviously quantum mechanical hamiltonians. However, in many cases, different choices of the hamiltonian describe the same physical model, in particular if the time variable is restricted to integral time values (in some units). On the other hand, one would like to implement \emph{locality} of a deterministic system, such as a cellular automaton, by postulating a hamiltonian that can be regarded as the integral (or sum) of a local hamilton density. This latter procedure is not straightforward. We found interesting examples where this can be done, but a systematic formalism still seems to be lacking.

Recently, we elaborated a scheme that enables us to map systems with pairs of integers onto other systems where these pairs are replaced by pairs of real but non commuting numbers\cite{GtH1}. This mapping requires the machinery of quantum mechanics. States of the discrete system may well be postulated to evolve completely deterministically, but this is no impediment against treating them as states in a Hilbert space. After our mapping is applied, much of their deterministic nature becomes hidden, and the system takes the shape of a quantum theory. Indeed, it is a real quantum theory, including Born's interpretation of the squares of the amplitudes as representing probabilities. Consequently, we arrive at a theory of quantum wave functions (which is what quantum states look like when defined on the real line), evolving completely in line with an ordinary Schr\"odinger equation, yet it is a deterministic theory in disguise. The importance of this mapping is its potential usefulness in quantum field theories. There, we have hamiltonians that are integrals of hamilton densities. We set out to follow the following strategy: turn a simple, exactly soluble cellular automaton into an exactly soluble quantum field theory, derive its hamiltonian, and subsequently add interaction terms. The hamiltonian will receive small, local corrections so that now a cellular automaton can be mapped onto an interacting quantum field theory. From there, find our way to the Standard Model.

We claim that,  in spite of its quantum mechanical appearance, the Standard Model may also be a deterministic system in disguise. The author is aware of the fact that most of his readers will not be prepared to jump so such a conclusion; in particular those who have Bell's inequaities\cite{Bell}\cite{epr}  in mind will not be inclined to accept the idea. But then we would invite the reader to study the transformation presented in this paper. Here, we display  how the mapping goes for non-interacting, massless bosons in one space- and one time dimension. The resulting quantum field theory is restricted to a discrete space-time lattice, but the lattice length, \(a\), may be taken to be as small as we wish, and indeed imposing a lattice cut-off is standard practice in quantum field theory. In more realistic theories, such a lattice is suspected to replace some much more sophisticated ultraviolet cut-off induced by the gravitational force at the Planck scale.

The quantum hamiltonian of the theory is modified by the lattice artifacts in a way to be specified later; in any case, there will be left-moving particles and right-moving particles that show no interactions at all. Being restricted to wave functions on the lattice, we see right-away that these particles represent quantised units of information moving to the left and to the right, both with the speed of light.

Our cellular automaton behaves just like this. It is defined by having two sets of integers: \(Q(x,t)\) are defined on the same lattice sites, while \(P(x,t)\) are defined on \(x\) integer but \(t\) is integer plus \(\half\). These  \(Q\) and \(P\) variables will now also be seen to send discrete packages of data to the left and to the right, just like our bosonic model, and for this reason one could already suspect that a mapping between the automaton and the bosonic model should exist. This is the mapping  that will be constructed in this paper. The mapping is far from trivial, as the reader will surely discover.

Initially, we intended also to describe the introduction of mass terms and interaction terms, along the lines sketched above, but in the 1+1 dimensional system we encountered an obstacle against this: both the cellular automaton and the quantum field theory feature a Goldstone symmetry\cite{Gold} forcing the excitations to be massless. Breaking the Goldstone symmetry would require non-local interactions of a type we would rather avoid, while it is known that interacting quantum field theories in 1+1 dimensions cannot have Goldstone symmetries\cite{Goldnointeract}. We conclude that the introduction of interactions and masses will require more advanced techniques or generalization to more dimensions, first.

Perhaps this is not necessary. Our mapping may have more direct implications for string and superstring theories. These do contain massless bosons on a 1+1 dimensional world sheet, so our observation may imply that determinism and discreteness may be considered in a systematic way for string theory. This could lead to new models, or at least new interpretations of the existing models.

\newsecl{The continuum quantum field theory}{QFT}

Consider first the standard, massless, bosonic field theory in 1+1 dimensions, described by a lagrangian and a hamiltonian,
	\be\LL=-\half(\pa_t q^2-\pa_x q^2)\ ;\qquad H=\int\dd x(\half p^2+\half\pa_x q^2)\ , \eel{LH}
where we use the symbol \( p(x)\) to denote the canonical momentum field associated to the scalar field \(q(x)\), which here obeys \( p(x)=\pa_t q(x)\).
The equal-time commutation rules are usually:
	\be[ q(x), q(y)]=[ p(x), p(y)]=0\ ;\qquad[ q(x), p(y)]=i\d(x-y)\ . \eel{standcomm}
If we now include the time variable by writing \( q(x,t)\) and \( p(x,t)\) in Heisenberg notation, the solution of the field equations can be written as follows:
	\be a^L(x,t)&=& p(x,t)+\pa_x q(x,t)\iss a^L(x+t)\ ;\crl{leftcont}
	 a^R(x,t)&=& p(x,t)-\pa_x q(x,t)\iss a^R(x-t)\ .\eel{rightcont}
In terms of these variables, the hamiltonian is
	\be H=\int\dd x\,\quart\bigg((a^L(x))^2+(a^R(x))^2\bigg)\ , \eel{HaLaRcont}
and using \(k\) to indicate momentum variables, we have in momentum space
	\be &&H=\int_0^\infty\dd k\,\half\bigg( (a^{L\dag}(-k)a^L(-k)+  a^{R\dag}(k)a^R(k)\bigg)\ , \crl{HaLaRk}
	k,k'>0:	&& [a^L(-k),a^L(-k')]=0\ ,\qquad [a^L(-k),a^{L\dag}(-k')]=2k\d(k-k')\ . \crl{aLaLdagcont}
		&& [a^R(k),a^R(k')]=0\ ,\qquad [a^R(k),a^{R\dag}(k')]=2k\d(k-k')\ . \eel{aRaRdagcont}
In our notation, \(a^{L,R}(k)\) are the annihilation and creation operators apart from a factor \(\sqrt{2k}\), so the hamiltonian~\eqn{HaLaRcont} can be written as
	\be H=\int_0^\infty\dd k\,(kN^L(-k)+kN^R(k))\ . \eel{contoccnumbers}
where \(N^{L,R}(\mp k)\dd k\) are the occupation numbers counting the left and right moving particles (we subtracted the vacuum values for the energy). 
The energies of these particles are equal to the absolute values of their momentum. All of this is completely standard and can be found in all the text books.

\newsecl{Notation}{notation} 
	In the previous Section, we still used (more or less) conventional notation, since everything was in the continuum. On the lattice, however, we will pick a somewhat less standard convention, as it will simplify the calculations there significantly. In the sequel, we will use 
		\be\hbox{capital Latin letters,}\quad &N,\  P,\  Q,\  X,\  \cdots,\!\! &\hbox{ to indicate integer valued fields,}\qquad\quad\crl{integers}
		\hbox{lower case Latin letters,}\, \quad&  a,\ p,\ q,\ \cdots, &\hbox{ to indicate  real numbers, and}\qquad\crl{real}
		\hbox{lower case Greek letters,}\quad &  \a,\ \eta,\  \xi,\  \l,\  \cdots, &\hbox{ for fractional numbers,} \eel{Greek} 
the latter being usually confined to the interval \((-\half,\ \half\,] \). An exception is the lattice coordinates \(x,\,t\), which are in lower case because they are not used as field variables.

Frequent use will be made of the number
		\be  \E=e^{2\pi}\approx\hbox{ 535{\small .49}}\,\cdots\ ,\ \hbox{ so that }\ 
		e^{2\pi i \a}\equiv\E^{i \a}\ ,\quad\hbox{and }\ \E^{iZ}=1\quad \hbox{if}\quad Z\in\Bbb Z\ . \eel{Exp}
If we have integers \(Q,\ P,\,\cdots\), we will often associate a Hilbert space of \emph{states} to these: \(|\,Q,\,P,\,\cdots\ket\). Then, there will be operators \(\eta_Q,\,\eta_P,\,\cdots\), defined by
	\be \E^{\ds iN\eta_Q}|\,Q,\,P,\,\cdots\ket=|\,Q+N,\,P,\,\cdots\ket\ ;\qquad  \E^{\ds iN\eta_P}|\,Q,\,P,\,\cdots\ket=|\,Q,\,P+N,\,\cdots\ket\ .\quad  \eel{etadef} 
In general, these operators will have eigenvalues restricted to be in the interval \((-\half,\,\half\,]\). 

The matrix elements of the \(\eta_Q\) operators can readily be calculated\cite{GtH1}: 
\be \eta_Q = \sum_{N\ne 0}{i\over 2\pi N}(-1)^N\E^{iN\eta_Q}\ ,\qquad \bra Q_1|\eta_Q|Q_2\ket={i\over 2\pi}(\d_{Q_1Q_2}-1){(-1)^{Q_2-Q_1}\over Q_2-Q_1}\ . \eel {etaQelements}
We will need to know the commutator between the operators \(\eta_Q\) and \(Q\):
	\be\bra Q_1|[\eta_Q,\,Q]|Q_2\ket =(Q_2-Q_1)\bra Q_1|\eta_Q|Q_2\ket={i\over 2\pi}\left(\d_{Q_1Q_2}-  (-1)^{Q_2-Q_1}\right)\ .\eel{etaQcomm}

Because we wish to keep factors \(2\pi\) in the exponents (to be absorbed when we use \(\E\) instead of \(e\)), our commutation rules for \(p\) and \(q\) operators (which are real numbers) will be
	\be[q,\,p]=i/2\pi\ , \eel{xpcomm}
which means that it is \(h\) rather than \(\hbar\) that we normalize to one.

\newsecl{Quantum field theory on the 1+1 dimensional lattice}{qftlattice}
Inserting a lattice cut-off for the UV divergences in quantum field theories is also standard practice. Restricting ourselves to integral values of the \(x\) coordinate, 
we replace the commutation rules ~\eqn{standcomm} by
	\be[ q(x), q(y)]=[   p^+  (x),   p^+  (y)]=0\ ;\qquad[ q(x),   p^+  (y)]=\fract i{2\pi}\d_{x,y}\eel{latticecomm}
(the reason for the superscript + will be explained later, Eqs.~\eqn{pdef} and \eqn{ppm}).
The exact form of the hamiltonian on the lattice depends on how we wish to deal with the lattice artifacts. The choices made below might seem somewhat artificial or special, but it can be verified that most alternative choices one can think of can be transformed to these by simple latticle field transformations, so not much generality is lost. It is important however that we wish to keep the expression~\eqn{contoccnumbers} for the hamiltonian: also on the lattice, we wish to keep the same dispersion law as in the continuum, so that all excitations must move left or right exactly with the same speed of light.
	
The lattice expression for the left- and right movers will be
	\be a^L(x+t)&=&   p^+  (x,t)+ q(x,t)- q(x-1,t)\ ;\crl{aL}
	     a^R(x-t)&=&   p^+  (x,t)+ q(x,t)- q(x+1,t)\ . \eel{aR} 
 They obey the commutation rules 
	    \be[a^L,a^R]=0\ ;&& [a^L(x),\,a^L(y)]=\pm\fract i{2\pi}\ \hbox{ if }\ y=x\pm 1\ ;\quad\hbox{else }\ 0\ ; \crl{aLaLcomm}
				&& [a^R(x),\,a^R(y)]=\mp\fract i{2\pi}\ \hbox{ if }\ y=x\pm 1\ ;\quad\hbox{else }\ 0\ . \eel{aRaRcomm}
In momentum space, writing
	\be a^{L,R}(x)\equiv\int_{-\half}^\half\dd\k\, a^{L,R}(\k)\,\E^{i\k x}\ , \eel{fouriera}
the commutation rules are
	 \be  [a^L(\k_1),\,a^L(-\k_2)]&=&-\fract 1\pi \,\d(\k_1-\k_2)\sin(2\pi\k_1)\ ; \qquad\crl{aLkcomm} 
      			 [a^R(\k_1),\,a^R(-\k_2)]&=&\fract 1\pi\, \d(\k_1-\k_2)\sin(2\pi\k_1)\ . \qquad\eel{aRkcomm} 

If we want the hamiltonian to take the form~\eqn{contoccnumbers}, 
then, in terms of the creation and annihilation operators \eqn{aLkcomm} and \eqn{aRkcomm},the hamiltonian must be
	\be H=\int_0^\half\dd\k{\pi\,\k\over \sin(2\pi\k)}\bigg(a^L(\k)\,a^L(-\k)+a^R(\k)\,a^R(-\k)\bigg)\ . \eel{hamiltonian}
Since, in momentum space, Eqs.~\eqn{aL} and \eqn{aR} take the form
	\be a^L(\k)=   p^+  (\k)+(1-\E^{-i\k}) q(\k)\ ,\qquad  a^R(\k)=   p^+  (\k)+(1-\E^{i\k}) q(\k)\ , \eel{aLRk}	
after some shuffling, we find the hamiltonian
	\be H=\pi\,\int_0^\half\dd\k\,\bigg({\k\over\tan\pi\k}\ |   p^+  (\k) |^2\ +\ 4\k\tan\pi\k\ | q(\k)+\half    p^+  (\k) |^2\bigg)\ , \eel{hampf} 
where \(|   p^+  (\k)|^2\) stands for \(   p^+  (\k)\,   p^+  (-\k)\). 
	Since the field redefinition \( q(x)+\half   p^+  (x)\ra q(x)\) does not affect the commutation rules, and
	\be \lim_{\k\ra 0}\,{\pi\k\over\tan(\pi\k)}=1\ ,\qquad 4\sin^2(\pi\k)| q(\k)|^2\ra|(\pa_x q)(\k)|^2\ , \eel{limtan}
 we see that the continuum limit~\eqn{LH}, \eqn{HaLaRk}  is obtained when the lattice length scale is sent to zero.

We now notice that the operators \(a^L(x,t)=a^L(x+t)\) and \(a^R(x,t)=a^R(x-t)\) move exactly one position after one unit time step. Therefore,
	\be a^L(x,1)=a^L(x+1,0)\ ,\qquad a^R(x,1)=a^R(x-1,0)\ ,\quad\hbox{etc.}\eel{timesteps}
and now we can use this to eliminate \(   p^+  (x,t)\) and \( q(x,t)\) from these equations. Writing
	\be    p^+  (x,t)\equiv p(x,t+\half)\ , \eel{pdef}
one arrives at the equations
	\be  q(x,t+1)&=& q(x,t)\ +\ p(x,t+\half)\ ; \crl{phievolve}
	p(x,t+\half)&=&p(x,t-\half)\ \,+\,  q(x-1,t)-2 q(x,t)+ q(x+1,t)\ . \eel{pevolve}
We now see why we had to shift the field \( q(x,t)\) by half the field momentum in Eq.~\eqn{hampf}:  it puts the field at the same position \(t+\half\) as the momentum variable \(   p^+  (x,t)\).

Thus, we end up with a quantum field theory where not only space but also time is on a lattice. The momentum values \(p(x,t+\half)\) can be viewed as variables on the timelike links of the lattice.
	
At small values of \(\k\), the hamiltonian~\eqn{hamiltonian}, \eqn{hampf} closely approaches that of the continuum theory, and so it obeys locality conditions there. For this reason, the model would be interesting indeed, if this is what the cellular automaton is going to match. However, there is a problem with it. At values of \(\k\) approaching \(\k\ra\pm\half\), the kernels diverge. Suppose we would like to write the expression~\eqn{hamiltonian} in position space as
	\be H=\sum_{x,\,s}M_{|s|}\,\bigg(a^L(x)\,a^L(x+s)+a^R(x)\,a^R(x+s)\bigg)\ , \eel{kernel}
then \(M_s\) would be obtained by fourier transforming the coefficient \(\pi\k/\sin(2\pi\k)\) on the unit interval for \(\k\). One obtains
	\be M_s=\int_{-\half+\l}^{\half-\l}{\k\dd\k\over\sin(2\pi\k)}\,\E^{- i s\k}\ =\fract 1{2\pi}\left\{\matrix{\log{2\over\l}-\sum_{k=0}^{s/2-1}{1\over k+1/2}\ \hbox{ if }\ s= 
		 \hbox{ even}\nm \\[2pt] \log( 2\l)+\sum_{k=1}^{(s-1)/2}{1\over k}\ \hbox{ if }\ s=\hbox{ odd}}\right. \eel{fourierham}
where \(\l\) is a tiny cut-off parameter. The divergent part is
	\be\fract 1{2\pi}\big(\log{1\over\l}\big)\,\sum_{x,y}(-1)^{x-y}\bigg(a^L(x)a^L(y)+a^R(x)a^R(y)\bigg)&=&\nm\\
	\fract 1{2\pi}\big(\log{1\over\l}\big)\bigg(\bigg(\sum_x(-1)^xa^L(x)\bigg)^2+\bigg(\sum_x(-1)^xa^R(x)\bigg)^2\bigg)\ .&&\eel{divpartham}

Also the kernel \(4\k\tan\pi\k\) in Eq.~\eqn{hampf} diverges as \(\k\ra\pm\half\). Keeping the divergence would make the hamiltonian non-local, as Eq.~\eqn{divpartham} shows. We can't just argue that the largest \(\k\) values require infinite energies to excite them because they do not; according to Eq.~\eqn{contoccnumbers}, the energies of excitations at momentum \(\k\) are merely proportional to \(\k\) itself.
We therefore propose to make a smooth cut-off, replacing the divergent kernels such as \(4\k\tan\pi\k\) by expressions such as
	\be (4\k\,\tan\pi\k)(1-e^{\L^2(\half-\k)^2})\ , \eel{momcutoff}
where \(\L\) can be taken to be arbitrarily large but not infinite.

\newsecl{The cellular automaton}{CA}
	Our cellular automaton is a model defined on a square lattice with one space dimension \(x\) and one time coordinate \(t\), where both \(x\) and \(t\) are restricted to be integers. The variables are two sets of integers, one set being integer numbers \(Q(x,t)\) defined on the lattice sites, and the other being defined on the links connecting the point \((x,t)\) with \((x,t+1)\). These will be called \(P(x,\,t+\half)\), but they may sometimes be indicated as 
		\be P^+(x,t)\equiv P^-(x,t+1)\equiv P(x,t+\half)\ . \eel{Ppmdef}
The automaton obeys the following time evolution laws:
		\be Q(x,t+1)&=&Q(x,t)\ +\ P(x,t+\half)\ ; \crl{Qevolve}
	P(x,t+\half)&=&P(x,t-\half)\ +\ Q(x-1,t)-2Q(x,t)+Q(x+1,t)\ , \eel{Pevolve}
just analogously to Eqs.~\eqn{phievolve} and \eqn{pevolve}.
It is also a discrete version of a \emph{classical} field theory where \(Q(x,t)\) are the field variables and \(P(x,t)={\pa\over\pa t}Q(x,t)\) are the \emph{classical} field momenta.

Imagining the \(x,t\) lattice to be very large, one might be interested in studying the statistical properties of the \(Q\) and \(P\) fields. To this end, we introduce Hilbert space, just as a tool. The basis elements of this Hilbert space are the \emph{states} \(\bigg|\,\{Q(x,0)\},\{P^+(x,0)\}\bigg\ket\).
If we consider a superposition of such states, we will simply \emph{define} the squares of the amplitudes to represent the probabilities. The total probability is  the length-squared of the vector, which will usually be taken to be one. At this stage, superpositions mean \emph{nothing} more than this, and it is obvious that any chosen superposition, whose total length is one, may represent a reasonable set of probabilities. The basis elements all evolve in terms of a permutation operator that permutes the basis elements in accordance with the evolution equations~\eqn{Qevolve} and \eqn{Pevolve}. As a matrix in Hilbert space, this permutation operator only contains ones and zeros, and it is trivial to ascertain that statistical distributions, written as ``quantum" superpositions, evolve with the same evolution matrix.

Indeed, in what follows later, we will make frequent use of operators in this Hilbert space, such as the operators \(\eta_Q(x)\) and \(\eta_P^+(x)\), which are defined exactly as in Eq.~\eqn{etadef} in Section~\ref{notation}, but now at each point \(x\) at time \(t=0\) (the operator \(\eta_P^+(x)\) acts on the integer \(P^+(x,0)\)).

The time variable \(t\) is an integer, so what our evolution equations generate is an operator \(U(t)\) obeying \(U(t_1+t_2)=U(t_1)U(t_2)\), but only for integral time.
In principle, it would be easy to find an operator \(H\), to be called ``hamiltonian", such that
	\be U(t)=\E^{\ds - i Ht}\ ;\qquad H=\sum_{n=1}^\infty{ (-1)^{n-1}i\over 2\pi n}(U(n)-U(-n))\ . \eel{hamiltonevolve}
This equation is obtained by Fourier analysis: when \(-\half<x<\half\), we have
	\be x=\sum_{n=1}^\infty a_n\sin 2\pi n x\ ;\qquad a_n\iss 2\int_{-\half}^\half  x\sin 2\pi x\iss {2(-1)^{n-1}\over 2\pi n}\ . \eel{fourierx}

The problem with this hamiltonian is that
	\bi{1.} It is not unique: one may add any integer to any of its eigenvalues; and
	\itm{2.} It is not extensive: if two parts of a system are spacelike separated, we would like the hamiltonian to be the sum of the two separate hamiltonians, but then it will quickly take values more than \(\halff\), whereas, by construction, the hamiltonian~\eqn{hamiltonevolve} will obey \(|H|\le \halff\). \ei
Thus, by adding appropriate multiples of integers to its eigenvalues, we would like to transform our hamiltonian into an extensive one. The question is how to do this.

Indeed, this is the question that forced us to do the investigations described in this paper; the hamiltonian of the quantum field theory considered here is an extensive one, and also naturally bounded from below. 

At first sight, however, the similarity between this automaton and the quantum field theory of Section~\eqn{qftlattice} may seem to be superficial at best. Quantum physicists will insist that the quantum theory is fundamentally different.

Our procedure will  force us first to compare the left-movers and the right-movers in both theories. In Appendix~\ref{movers}, some of the preperties of systems obeying the evolution equations~\eqn{Qevolve} and \eqn{Pevolve} are listed. Here also, it is found that the operators \(a^L(x)\) and \(a^R(x)\) have the desirable properties that they are left movers and right movers, and furthermore, they can directly be expressed in terms of the original field variables. In the automaton, these are integers, so, the results of Appendix~\ref{movers} imply that we must write them as:
	\be A^L(x+t)&=&P^+(x,t)+Q(x,t)-Q(x-1,t)\ ;\crl{ALdef}
	     A^R(x-t)&=&P^+(x,t)+Q(x,t)-Q(x+1,t)\ . \eel{ARdef}  
Note, that these are fields that transport streams of data to the left or to the right, but at this point the resemblance to elementary particles may still seem to be remote. 

\newsecl{The mapping}{map}
	We now claim that all states of the cellular automaton of Section~\ref{CA} can be mapped onto the quantum field theory states of Section~\ref{QFT}, with a few exceptions: the mapping is nearly though not quite one-to-one. Several ways to describe this mapping were tried, and we found the most transparent one to be as follows. We first consider the operators \(\eta_P(x)\) and \(\eta_Q(x)\), as described in Sections ~\ref{notation} and \ref{CA}. They will be identified with operators in the quantum field theory, by observing the following.
	
If we have operators \(q_i\) and \(p_i\) obeying the commutation rules
	\be [q_i,\,p_j]=\fract i{2\pi}\d_{ij}\ , \eel{pqcomm}
then we can readily derive the commutation properties of the exponentiated operators:
	\be 	\Ex{iap_i}\,\Ex{ibq_j}=\Ex{iab\,\d_{ij}}\,\Ex{ibq_j}\,\Ex{iap_i}\ . \eel{expcomm}
This means that, if \(a\) and \(b\) are both integers, \(\E^{iap_i}\) and \(\E^{ibq_j}\) commute. If we now define the operators \(\eta_{Qi}\) and \(\eta_{Pj}\) as follows,
	\be \E^{i\eta_{Qi}}=\E^{-ip_i}\ ,\qquad \E^{i\eta_{Pj}}=\E^{iq_j}\ , \eel{etapqdef}
while furthermore restricting \(\eta_{Qi}\) and \(\eta_{Pj}\) all to lie within the interval \((-\half\,\half\,]\), then they are uniquely defined, and they all commute. Thus, we can define all operators \(\eta_Q(x)\) and \(\eta_P^+(x)\) of our cellular automaton as (minus) the fractional parts of the field momentum operators \(   p^+  (x,0)\) and the field operators \( q(x,0)\) of our quantum field theory. Note, that \(\E^{-i   p^+  (x)}\) raises the field \( q(x)\) by one unit and \(\E^{i q(x)}\) raises the field \(   p^+  (x)\) by one unit, so indeed the operators that raise or lower fields \( q(x,0)\) by one unit, will now do the same for the variable \(Q(x,0)\) of the automaton, and the operators the raise or lower \(   p^+  (x,0)\) in the field theory, do the same for the variables \(P^+(x,0)\) in the cellular automaton. In the automaton, these raising and lowering operators all commute, but so they do in the quantum field theory.	
	
Note, that the identification of the raising and lowering operators in the automaton with the ones in the quantum field theory, is restricted to be valid for the states at time \(t=0\). To find how the identification is done at other times, we use the Heisenberg notation: \emph{operators} are time dependent, while \emph{states} are time independent.	

What has been achieved so-far is that \emph{if} we have identified \emph{one} state in the cellular automaton with \emph{one} state in the quantum field theory, then we can identify all other states, simply by applying the raising and lowering operators as often as we want\fn{Indeed, this part of the identification is valid in many more cellular automaton models that can be identified with quantum field theories, also in higher dimensions; it is the next part of the argument however, that we do not know exactly how to generalize.}. How do we identify one state? We can rephrase this question: 

\noindent\emph{We identified the fractional parts of the fields \( q(x,t)\) and \(   p^+  (x,t)\) with cellular automaton operators. How do we identify the integral parts? }

We can't just identify them with the variables \(Q(x,t)\) and \(P(x,t)\) because the time evolution laws will round off the fractional parts differently.  And therefore, this turns out to be not so easy. Take the cellular automaton ``vacuum" state 
	\be |\W\ket\equiv|\{Q(x,0)=0,\ P^+(x,0)=0\}\ket\ . \eel{zerostate}
Could this be mapped onto the ground state of the quantum field theory's hamiltonian? This is probably incorrect since the state \(|\W\ket\) is orthogonal to all states obtained by raising or lowering one of its \(Q\) or \(P^+\) values, but when applying one of the operators \(\E^{iN\,p^+(x,0)}\) or  \(\E^{iN q(x,0)}\) to the energy ground state of the quantum field theory,  these in general do \emph{not} annihilate this state.

Could \(|\W\ket\) be mapped onto the state \(\prod_x\j_0( q(x,0))\), where \(\j_0\) is the wave function derived in ref~\cite{GtH1}, which \emph{is} orthogonal to all those states that would be obtained by raising or lowering \(Q\) and/or \(P^+\) values? Again, this would probably be wrong, because we cannot prove that this state is time-independent (even when we restrict ourselves to integer time values), whereas the state \(|\W\ket\) is.

The correct mapping is obtained by viewing Hilbert space as the product of the space spanned by the left-movers and that spanned by the right-movers. 
We do the mapping for the left-movers and the mapping for the right-movers separately.

So let us first concentrate on the left movers. The cellular automaton is then described by the integers \(A^L(x+t)\), as defined in Eq.~\eqn{ALdef}. In that Hilbert space, we also have the operators \(\eta_A(x+t)\), defined by
		\be \E^{iN\eta_A(x)}|\{A(y)\}\ket=|\{A'(y)=A(y)+N\d_{xy}\}\ket\ , \eel{etaAdef}
where the superscript \(L\) was temporarily omitted. 
As in Eq.~\eqn{etaQelements}, we can write the matrix elements of \(\eta_A\):
	\be \bra A_1(x)|\eta_A(x)|A_2(x)\ket={i\over 2\pi}(\d_{A_1(x),A_2(x)}-1){(-1)^{A_2(x)-A_1(x)}\over A_2(x)-A_1(x)}\ , \eel {etaAelements}
while \(\eta_A(x)\) commutes with all operators \(A(y)\) and \(\eta_A(y)\) when \(y\ne x\).

From Eq.~\eqn{etaAelements}, we derive the commutator between \(\eta_A\) and \(A\). This is as in Eq.~\eqn{etaQcomm}:
	 \be\bra A_1|[\eta_A,\,A]|A_2\ket ={i\over 2\pi}\left(\d_{A_1A_2}-  (-1)^{A_2-A_1}\right)\ .\eel{etaAcomm}
This may be written as
	\be[\eta_A,\,A]=\fract i{2\pi}({\Bbb{I}}-|\j_A\ket\bra\j_A|)\ ;\qquad\bra A|\j_A\ket\equiv(-1)^A\ . \eel{commedge1}
\(|\j_A\ket\) is a very special state that is defined at every spacelike point \(x\). So we have
	\be[\eta_A(x),\,A(y)]=\fract i{2\pi} \d_{xy}(\Bbb I-|\j_A(x)\ket\bra\j_A(x)|)\ . \eel{etaAAxcomm}
We call \(|\j_A(x)\ket\) an \emph{edge state}, since it lives on the edge of the interval in \(\eta\) space: \(\eta_A(x)=\pm\half\). 

Our next attempt to construct the mapping consisted of omitting these bothersome edge states. Then,
	\be [\eta_A(x),\,A(y)]\qu\fract i{2\pi}\,\d_{xy}\  . \eel{regcomm}
This would allow us construct operators \(a^L(x)\) that obey the commutation rules of the quantum field theory, Eq.~\eqn{aLaLcomm}. Re-inserting the superscript \(L\):
	\be a^L(x) \qu A^L(x)+\eta_A^L(x-1)\ . \eel{aALidentificationattempt}
and similarly we coud define the right-movers. The good point about this attempt is, that the time evolution just shifts these operators one step to the left or to the right, without mixing them. Since in the cellular automaton, \(A^L(x)\) and \(\eta_A^L(x)\) shift the same way, this mapping commutes with the time evolution.
	
Eq.~\eqn{aALidentificationattempt} is not demonstrably false, since the constraint that we limit ourselves to the states that are orthogonal to the edge states \(|\j_A(x)\ket\) does commute with the time evolution. However, it is an unnatural constraint. The edge states would have finite energy, in general, so using these operators \(a^L\) and \(a^R\) as creation and annihilation operators might lead to wrong results.

A superior mapping procedure is the one introduced in Ref.~\cite{GtH1}. We know how the \(\eta\) operators are transformed, but to express the operators \(P^+\) and \(Q\) themselves, or equivalently, the operators \(A^L(x)\) and \(a^R(x)\), we have to establish a phase function in \(\eta\) space, which is now an infinite-dimensional torus, that is, the product of all circles defined by the periodic operators \(\eta_P^+(x,0)\) and \(\eta_Q(x,0)\). The procedure goes exactly as in Ref.~\cite{GtH1}.

We replace \eqn{aALidentificationattempt} by
	\be a^L(x)\iss -\fract i{2\pi}\,{\pa\over\pa\eta_A^L(x)}\ +\ {\pa\over\pa\eta_A^L(x)}\,\vv({\vec\eta_A^L(x)})\ -\ \eta_A^L(x-1)\ , \eel{aALidentification}
where \(\vv({\vec\eta_A^L})\) is a phase function that depends on all values of \(\eta_A^L(x)\), but it is the same function for all points \(x\). One easily derives that, indeed the commutation relation \eqn{aLaLcomm} holds --- as long as \(\vv\) is non-singular, and the term \(\eta_A^L(x-1)\) should not jump. 

But, of course, \(\eta_A^L(x-1)\) does jump, when we restrict it to stay within the interval \((-\half,\,\half\,]\). This is what we need the phase \(\vv\) for. It is postulated to obey the periodicity conditions
	\be\vv(\{\eta_A^L(x) +\d_{x,x_1}\})\ =\ \vv(\{\eta_A^L(x)\})+\eta_A^L(x_1+1)\ . \eel{vhiperiods}
If, in Eq.~\eqn{aALidentification}, we substitute \(x-1=x_1\), we see that this expression is now strictly periodic in \(\eta\) space, so now we expect no contribution from edge states.

Yet there is a contribution from an edge state, and this is because the periodicity condition~\eqn{vhiperiods} cannot hold everywhere, since there is a clash. If we go around the torus in two different directions, to return to the same point, a full phase rotation results. Thus \(\vv(\{\eta_A^L(x)\})\) has vortices trapped in it, and these cause new singularities.

The function is easy to construct \cite{GtH1}. Write
	\be\vv(\{\eta_A^L\})=\sum_x\vv(\eta_A^L(x+1),\eta_A^L(x))\ ;\qquad r(\eta,\xi)\E^{i\vv(\eta,\xi)}\equiv\sum_{K=-\infty}^\infty\E^{-\half(K-\xi)^2-Ki\eta}\ , \eel{vhidef} where \(r\) and \(\vv\) are both real functions of \(\eta\) and \(\xi\).
		This is a special case of the elliptic theta function \(\vartheta_3\), and it can also be written as a product\cite{GR} instead of a sum\fn{The function \(r(\eta,\xi)\) here differs by a factor \(\E^{-\half\xi^2}\) from the function used in Ref\cite{GtHCA}.}. Due to this property, one can also write the function \(\vv(\eta,\xi)\) as follows:
			\be\vv(\eta,\xi)=\sum_{K= 0}^\infty\bigg(\arg(1+\E^{i\eta+\xi-K-\half})+\arg(1+\E^{-i\eta-\xi-K-\half})\bigg)\ . \eel{vhiarg}
There is a duality property:
	\be \vv(\eta,\xi)+\vv(\xi,\eta)=\eta\,\xi\ . \eel{vhiduality}
Eq.~\eqn{aALidentification} simplifies into
	\be a^L(x)=-\fract i{2\pi}\,{\pa\over\pa\eta_A^L(x)}\ +\ {\pa\over\pa\eta_A^L(x)}\bigg(\vv(\eta_A^L(x+1),\eta_A^L(x))-\vv(\eta_A^L(x-1),\eta_A^L(x))\bigg)\ . \eel{aALsimplification}

We can now compute\cite{GtH1} the matrix elements of \(a^L(x)\) when sandwitched between two cellular automaton states \(|\{A^L(x)\}\ket\):
	\be\bra\vec{A_1^L}|a^L(x)|\vec{A_2^L}\ket \iss A^L(x)\d&+&\d\bigg|_{x,x+1}{(-1)^{\D A^L(x)+\D A^L(x+1)+1}i\D A^L(x+1)\over \D A^L(x)^2+\D A^L(x+1)^2} 		\nm\\[2pt] &-&\ \d\bigg|_{x,x-1}{(-1)^{\D A^L(x)+\D A^L(x-1)+1}i\D A^L(x-1)\over \D A^L(x)^2+\D A^L(x-1)^2}\ ,	\quad	\eel{aLmatrix}
where \(\d\) is the identity matrix in \(\vec {A^L}\) space (so that \(A_1^L(x)=A_2^L(x)=A^L(x)\)), while \(\d|_{y,z}\) stands for the matrix that keeps all values of 
\(A_2^L(x)\) equal to \(A_1^L(x)\), except at the points \(x=y\) and \(x=z\), where we define
	\be \D A^L(y)\equiv A_2^L(y)-A^L_1(y)\ , \qquad \D A^L(z)\equiv =A_2^L(z)-A^L_1(z)\ ; \eel{deltadef}
(the diagonal element of \(\d|_{y,z}\), where the terms in \eqn{aLmatrix} would be ill-defined, are taken to be zero).

It is now important to check the commutation rules for the field \(a^L(x)\). Since the operators \(A^L(x)\) refer to the states of the automaton, they all commute with one another, and since the additional terms in \eqn{aLmatrix} are diagonal in \(\eta\) space, they also commute with one another (this is trivially seen to be true in Eq.~\eqn{aLmatrix} because the matrix elements only depend on \(\D A^L\), not on the \(A_{1,2}^L(x)\) themselves). The non-commuting parts come from the cross terms. Since a matrix element of \([X,\,A^L(x)]\) in \(A^L\) space is the matrix element of \(X\) multiplied by \(\D A^L(x)\), one finds

 	\be [a^L(x),\,a^L(x+1)]=\fract i{2\pi}\bigg(1-|\j_A^L(x,x+1)\ket\bra\j_A^L(x,x+1)|\bigg)\ , \eel{commedge}
where the new edge state is defined by
	\be\bra A^L(x),\,A^L(y)\,|\,\j_A^L(x,y)\ket=(-1)^{A^L(x)+A^L(y)}\ . \eel{newedge}
This edge state originates from the flux singularity in the phase function \(\vv(\eta,\xi)\) located at the corners \(\eta=\pm\half,\,\xi=\pm\half\), and it can be replaced but not removed. We claim that this state is far less harmful than the edge states encountered previously.  The point is that any state that is not orthogonal to any of these edge states, carries infinite energy. This we see by calculating the expectation value of the hamiltonian \eqn{kernel}. Plugging the matrix elements \eqn{aLmatrix} of \(a^L(x)\) in here, we see a logarithmic divergence of positive terms (when \(s=0\)), and negative terms (when \(s=2\)), but the positive ones dominate since \(M_0>M_2\). The odd \(s\) terms do not diverge.

We conclude that the edge state contribution in Eq.~\eqn{commedge} can be neglected whenever states with finite energy are considered. In our construction, all eigen states of the operators \(a^L(x)\) are unique. Repeating the same procedure for the \(a^R\) operators gives us a complete mapping of all finite energy states of the quantum field theory onto states of the cellular automaton. The mapping commutes with the time evolution, since both the \(a^L(x)\) operators and the \(a^R(x)\) operators, just like the ones of the cellular automaton, merely shift to the left or to the right as time evolves. An important restriction is that we should only look at those elements of the automaton states that are orthogonal to the edge states:
	\be\sum_{P,Q}(-1)^{P+Q}\bra P,Q|\j\ket=0\ , \eel{edgecondition}
where \((P,Q)=(\,A^L(x),\,A^L(x+1)\,)\) or  \((P,Q)=(\,A^R(x),\,A^R(x+1)\,)\)  for any point \(x\) on the lattice.

Curiously, this condition excludes the pure basis elements of the automaton such as the state \(|\W\ket\), which is why in a more sophisticated theory such a constraint might be undesirable. We observe that the exceptional states, the edge states \eqn{newedge}, consist of  a single state at every site. As such, it may describe \emph{fermions} on the lattice. In a next publication, it will be explained that the bothersome constraint is lifted if we  map the cellular automaton of Section~\ref{CA} onto a \emph{supersymmetric} 1+1 dimensional quantum model, with one bosonic field together with one fermion.

In the quantum field theory, we found another mild restriction: all states with particles in the very highest energy mode, \(\k=\pm\half\), contribute non-locally to the hamiltonian, since the kernel \eqn{hamiltonian} diverges there. This is why we propose the cut-off \eqn{momcutoff}.

We emphasize once more: the important reason why this mapping is of interest is, that now we know the hamiltonian, Eq.~\eqn{kernel}, where we can plug in the operators \(a^{L,R}(x)\) since we know their matrix elements \eqn{aLmatrix}.

This result is quite non-trivial. Since this hamiltonian correctly transports the left movers to the left and the right movers to the right, by one step in every unit of time, also the integral parts \(A^L(x)	\) and \(A^R(x)\) move correctly, as well as their fractional parts, \(\eta_A^L(x-1)\) and \(\eta_A^R(x+1)\). So this is the correct hamiltonian that generates the evolution law of the automaton. Being identical to a quantum field theory hamiltonian, it no longer suffers from any positivity problem, which was an important difficulty in previous work\cite{GtHCA}.

\newsecl{Interactions}{interact}

Up till this point, our construction might be looked upon as a mere mathematical curiosity, but physically rather boring, since both the cellular automaton and the quantum field theory decribe nothing but non-interacting particles. The left-movers do not react at all on the presence of right-movers and \emph{vice versa}. 

In principle, introducing interactions at both sides of the mapping seems to be straightforward. Suppose that we make a minor modification in the cellular automaton's evolution law. After every move as the time clock counts, it is checked whether some rather improbable combination of cellular field variables occurs, for example: at any value of \(x\),
	\be Q(x)=0,\ P^+(x)=5,\  Q(x+1)=17,\ Q(x-1)=3\ . \eel{special}
Only if this situation occurs, \(P^+(x)\) is augmented by one unit. Another condition is considered where \(P^+(x)\) decreases by one unit. We write this action as a permutation matrix:
	\be {\mathcal{P}}\equiv\E^{-iH^\intt}\ ,\qquad H^\intt=\l\sum_x{\mathcal{H}}^\intt(x)\ . \eel{intham}
We already had the hamiltonian \(H_0=\sum_x{\mathcal{H}}_0(x)\), the hamiltonian~\eqn{kernel}. To write the effect of both interactions in a hamiltonian form, we use Baker-Campbell-Hausdorff\cite{BCH}:
	\be e^A\,e^B=e^{A+B+\half[A,B]+\cdots}\ , \eel{BCH}
where the dots represent an infinite series of commutators. The fact that these are commutators is very important, because it means that the net result can again be written as the integral over a hamiltonian density, and this means that locality of the theory is not dramatically affected. Now, we took the condition \eqn{special} to be sufficiently uncommon so that its effect may be expected to be small. Therefore, perturbation expansion in powers of the parameter \(\l\) in Eq.~\eqn{intham} makes sense; the higher terms only contribute of the condition is met in several ways, which will reduce the numerical values of those contributions even more. We do see now that the quantity \(B+\half[A,B]+\cdots\) in Eq.~\eqn{BCH} generates an `interaction hamiltonian' in the quantum field theory. It is important to observe that, with this modification, the hamiltonian not only represents a theory with interaction, but it still can be written as the sum over hamiltonian densities. Since the hamiltonian density is corrected by finite correction terms, it continues to have well-defined lower bounds: the vacuum state.

However, in this particular case, there is a serious complication that practically invalidates this procedure: to express the original fields \( q(x,t)\) of the cellular automaton in terms of the \(a^L\) and \(a^R\) operators that we constructed, one has to invert Eq.~\eqn{aLRk} , and for small \(\k\) values this blows up. Clearly, only the derivatives of the \(q\) fields occur in the original model, not the \(q\) fields themselves. So, in our special case, it seems that we can only handle interactions where the interaction hamiltonian is a function of \(\pa q/\pa x\) rather than \(q\) itself. However, if we try to get interactions this way, \emph{we cannot lift the Goldstone symmetry\cite{Gold} that made the particles massless}. Not only might our original intention  to create interaction and mass terms  be difficult to realize in this model, so that any attempts to create mass terms might give rise to unacceptable non-localities; unfortunately, this also jeopardises our attempts to get any interactions at all, because interacting field theories in 1+1 dimensions are known not to allow for a Goldstone realization of a continuous symmetry\cite{Goldnointeract}. We believe that this difficulty is an artifact of the model considered; perhaps more sophisticated models can be found, in particular  in higher space dimensions, where such difficulties do not arise. As for this paper, we do not expand any further on the topic of interactions.

Our main intention with this paper was to demonstrate in what way cellular automata can actually behave as quantum field theories and \emph{vice versa}.

\newsecl{Discussion}{disc} 

It is difficult to explain more clearly the theory that there may be determinism behind what is called quantum mechanics today, than by pointing to the mapping displayed in this paper. Every operator in the quantised field theory has its counterpart in the variables describing a cellular automaton. What exactly this means concerning the interpretation of the EPR paradox\cite{epr} and Bell's inequalities\cite{Bell}, is difficult to phrase precisely. An attempt was made by the author in Ref.\cite{GtHcollapse}. For one thing, our theory is definitely not a hidden variable theory of the type that Bell considered in his paper\cite{Bell}; however these arguments clearly failed to convince many people. Perhaps the best argument consists of stating that our theory is nothing but an extrapolation of the mathematics phrased here, claiming that there is no logical obstacle against continuation of these models towards systems that more and more resemble the Standard Model of the Elementary Particles.

Yet there are a few subtleties. A subset of the cellular automaton states, being all states that are not orthogonal to the edge states, have to be left out of the mapping, as they would have infinite energy in the quantised field theory. This constraint is not insignificant, because it means that for all left moving modes and for all right moving modes, at every pair of points, Eq.~\eqn{edgecondition} implies that we must postulate that
	\be\sum_{A^L(x),\,A^L(x+1)}(-1)^{A^L(x)+A^L(x+1)}|\vec{A^L}\ket =0\ , \eel{edgeconstraint}
and the cellular ground state, \(|\W\ket=|0,0,\cdots\ket\), does not obey this constraint. We must conclude that either the quantum field theoretical description of cellular automata only applies to automata that are in rather highly excited modes, far away from the trivial state \(|\W\ket\), or we have to add \emph{fermions} to the bosonic theory.

Secondly, the resulting quantum field theory has a hamiltonian that approaches the continuum theory smoothly as the lattice length \(a\) tends to zero, but the lattice edge states, the particles at maximal momentum, \(\k=\pm\half\), contribute to the hamiltonian non-locally, so that the kernel \(M_s\) of the hamiltonian is non-local after all. We would insist that this feature would be unnoticeable for observers in a universe who have no access to particles with such a high momentum.

Note however that we do have all states of the quantised field theory in our cellular automaton, as we reproduce Eq.~\eqn{contoccnumbers}, describing all possible occupation numbers of the particles. Note also that the cellular automaton model itself is \emph{entirely local}, so if we use that as our ``ontological" underlying theory, the apparent signals of some non-locality in the effective bosonic theory cannot be used to explain what some observers regard as a necessary non-locality in hidden variables theories for quantum mechanics.

In principle, interactions can be introduced, by adding small modifications to the cellular evolution rule, as explained in Section~\ref{interact}. However, it is also explained there that there are pathologies in 1+1 dimensions. These are caused by the Goldstone symmetry that we built in. As is well-known\cite{Goldnointeract}, interacting quantum field theories cannot realise Goldstone symmetries in 1+1 dimensions, while we also found it difficult to break the symmetry. This also obstructs the generation of mass terms.

Our mapping may have interesting implications for string theory. Bosonic strings are harmonic theories on a one plus one dimensional string world sheet. If we introduce a lattice in the world sheet, strings become small solid `rods' stitched together, but in our model also target space, the 10 or 26 dimensional space-time, was made discrete. This means that our cellular automaton is relevant for string theory, describing discrete strings on a space-time lattice.\cite{Suss}

\appendix
 \newsecl{Left and right movers}{movers}
The cellular automaton as well as the quantised field theory will be postulated to obey the same equations on the lattice. In both systems, we locate the field variables   \( q(x,t)\)  on the regulare lattice sites \((x,t)\) where both \(x\) and \(t\) are restricted to be integers. The momentum variables are located conveniently on half-odd time coordinates, but sometimes it will be more practical to locate them on integer spots. Therefore, we write
	\be p^\pm(x,t)\equiv p(x,t\pm\half)\ . \eel{ppm}
While in the quantized field theory these variables are operators with a continuous spectrum of eigenvalues, obeying the usual commutation rules~\eqn{latticecomm}, the cellular automaton is described by replacing \( q(x,t)\ra Q(x,t)\) and \(p(x,t)\ra P(x,t)\), where \(Q\) and \(P\) only take integer values and all commute. In this Appendix we keep the notation \(q\) and \(p\).

The evolution equations are Eqs.~\eqn{phievolve} and \eqn{pevolve}.
It is sometimes useful to have the complete solution of these equations forward in time:
	\be q(0,t)&=&\sum_{n=1}^tp^+(2n-1-t,\ 0)+\sum_{m=1}^{2t-1}(-1)^{m-1} q(m-t,\ 0)\ ;\crl{phisol}
	p^+(0,t)&=&\sum_{m=0}^{2t}(-1)^m\bigg(p^+(m-t,\ 0)+2 q(m-t,\ 0)\bigg)\ -\  q(t,\ 0)- q(-t,\ 0)\ .\qquad\eel{psol}
The \(q\) and \(p\) fields can be split up in left-movers and right-movers:
	\be q(x,t)= q^L(x+t)+ q^R(x-t)\ ;\quad p(x,t+\half)=p^L(x+t+\half)+p^R(x-t-\half)\ ,\quad \eel{leftright}
which obey the equations
	\be p^L(x+\half)= q^L(x+1)- q^L(x)\ ;\qquad p^R(x-\half)= q^R(x-1)- q^R(x)\ . \eel{pphimovers}
However, we will often consider the more fundamental left and right movers \(a^L,\ a^R\), defined by inverting the equations~\eqn{leftright} :
	\be a^L(x)\iss p^L(x+\half)+p^L(x-\half)&=&p^+(x,0)+ q(x,0)- q(x-1,0)\ ;\crl{aLdef}
	     a^R(x)\iss p^R(x+\half)+p^R(x-\half)&=&p^+(x,0)+ q(x,0)- q(x+1,0)\ . \eel{aRdef}  
It is easy to verify from the basic equations~\eqn{phievolve} and \eqn{pevolve} that the rhs of equations~\eqn{aLdef} and \eqn{aRdef} will roll to the left and to the right, respectively.   In momentum space, \(a^L\) and \(a^R\) will be associated to the particle creation and annihilation operators of the field theory.

Assuming the commutation relations \eqn{latticecomm}, the operators \(a^L\) and \(a^R\) will obey easy commulation rules themselves, 
as given in Eqs.~\eqn{aLaLcomm} and \eqn{aRaRcomm}, while the commutation rules for \( q^L,\,\  q^R\), \(p^L\) and \(p^R\) will be non-local. This is why it would be difficult to introduce interactions that are expressed directly in terms of the \( q(x)\) and \(   p^+  (x)\) operators.

\end{document}